\documentclass[prb, twocolumn, superscriptaddress]{revtex4-2}
\usepackage{bm, amsmath, amsfonts, amssymb, mathtools, braket}
\usepackage{multirow}
\usepackage{graphicx}
\usepackage{float, color}

\usepackage[
pagebackref=false,
colorlinks=true,
linkcolor=blue,
urlcolor=blue,
filecolor=black,
citecolor=red,
pdfstartview=FitV,
pdftitle={},
pdfauthor={},
pdfsubject={},
pdfkeywords={},
pdfpagemode=None,
bookmarksopen=true
]{hyperref}

\begin{document}

\newcommand{\ii}{\text{i}}

\title{Many-body topology of non-Hermitian systems}

\author{Kohei Kawabata}
\email{kawabata@cat.phys.s.u-tokyo.ac.jp}
\affiliation{Department of Physics, University of Tokyo, 7-3-1 Hongo, Bunkyo-ku, Tokyo 113-0033, Japan}

\author{Ken Shiozaki}
\email{ken.shiozaki@yukawa.kyoto-u.ac.jp}
\affiliation{Yukawa Institute for Theoretical Physics, Kyoto University, Kyoto 606-8502, Japan}

\author{Shinsei Ryu}
\email{shinseir@princeton.edu}
\affiliation{Department of Physics, Princeton University, Princeton, New Jersey, 08540, USA}

\date{\today}

\begin{abstract}
Non-Hermiticity gives rise to unique topological phases that have no counterparts in Hermitian systems.
Such intrinsic non-Hermitian topological phases appear even in one dimension while no topological phases appear in one-dimensional Hermitian systems.
Despite the recent considerable interest, the intrinsic non-Hermitian topological phases have been mainly investigated in noninteracting systems described by band theory.
It has been unclear whether they survive or reduce in the presence of many-body interactions.
Here, we demonstrate that the intrinsic non-Hermitian topological phases in one dimension survive even in the presence of many-body interactions.
We formulate a many-body topological invariant by the winding of the complex-valued many-body spectrum in terms of a $\mathrm{U} \left( 1 \right)$ gauge field (magnetic flux).
As an illustrative example, we investigate the interacting Hatano-Nelson model and find a unique topological phase and skin effect induced by many-body interactions.
\end{abstract}

\maketitle

%%%%% Introduction %%%%%
\section{Introduction}

Recently, topological characterization of non-Hermitian systems has attracted considerable interest both in theory~\cite{Rudner-09, Sato-11, *Esaki-11, Hu-11, Schomerus-13, Longhi-15, Lee-16, Leykam-17, Xu-17, Xiong-18, Shen-18, *Kozii-17, MartinezAlvarez-18, Gong-18, YW-18-SSH, *YSW-18-Chern, Kunst-18, McDonald-18, Lee-Thomale-19, Liu-19, Lee-Li-Gong-19, Longhi-19, KSUS-19, ZL-19, Herviou-19, *Herviou-19-ES, Hirsbrunner-19, Zirnstein-19, *Zirnstein-21, Borgnia-19, KBS-19, Yokomizo-19, MCClarty-19, Okuma-19, Bergholtz-19, JYLee-19, Schomerus-20, Chang-20, Zhang-20, OKSS-20, Yi-Yang-20, KOS-20, Terrier-20, Bessho-20, Denner-21, Okugawa-20, KSS-20, Claes-21, KSR-21, Okuma-21, Sun-21, Okugawa-21, Vecsei-21, Shiozaki-21} and experiments~\cite{Poli-15, Zeuner-15, Zhen-15, Weimann-17, Xiao-17, St-Jean-17, Bahari-17, Zhao-18, Zhou-18, Harari-18, *Bandres-18, Zhao-19, Brandenbourger-19-skin-exp, *Ghatak-19-skin-exp, Helbig-19-skin-exp, *Hofmann-19-skin-exp, Xiao-19-skin-exp, Weidemann-20-skin-exp, Wang-21S, *Wang-21N, Wengang-21, Palacios-21, Zhang-21}.
Non-Hermiticity originates from exchanges of particles or energy with the external environment and gives rise to unique phenomena in open classical and quantum systems~\cite{Konotop-review, Christodoulides-review, BBK-review}.
The rich behavior of non-Hermitian topological systems is due to the complex-valued spectrum, which enables two types of complex-energy gaps: 
line and point gaps~\cite{KSUS-19}.
In the presence of a line gap, non-Hermitian systems can be continuously deformed to Hermitian systems. 
Thus, the line-gap topology describes the stability of Hermitian topology against non-Hermitian perturbations and is relevant to, for example, topological lasers~\cite{Schomerus-13, KSUS-19, Poli-15, St-Jean-17, Bahari-17, Zhao-18, Harari-18, *Bandres-18, Zhao-19}.
In the presence of a point gap, on the other hand, non-Hermitian systems can only be continuously deformed to unitary systems.
Consequently, the point-gap topology can describe topology that has no counterparts in Hermitian systems and is intrinsic to non-Hermitian systems.
For example, such intrinsic non-Hermitian topology appears in one dimension~\cite{Gong-18} while one-dimensional Hermitian systems cannot support topological phases without symmetry protection~\cite{HK-review, QZ-review, CTSR-review}.
The point-gap topology describes unique non-Hermitian topological phenomena such as the unidirectional dynamics~\cite{Longhi-15, Gong-18, McDonald-18, JYLee-19, KSR-21, Weidemann-20-skin-exp} and the emergence of exceptional points~\cite{Leykam-17, Xu-17, Shen-18, KBS-19, Zhen-15, Zhou-18}.
Furthermore, it is the topological origin of the non-Hermitian skin effect~\cite{Zhang-20, OKSS-20}, which is the anomalous localization induced by non-Hermiticity~\cite{Lee-16, YW-18-SSH, Kunst-18, Lee-Thomale-19, Yokomizo-19, Brandenbourger-19-skin-exp, *Ghatak-19-skin-exp, Helbig-19-skin-exp, *Hofmann-19-skin-exp, Xiao-19-skin-exp, Weidemann-20-skin-exp, Palacios-21, Zhang-21}.

Despite the considerable interest in non-Hermitian topological systems, much work has hitherto focused on noninteracting systems.
While several recent works investigated interacting non-Hermitian topological systems~\cite{Guo-20, Yoshida-19, *Yoshida-20, Mu-20, Xi-19, Lee-20, Matsumoto-21, Zhang-20Mott, Liu-20, Xu-20, Shackleton-20, CHLee-20, *Shen-21, Yang-21, Pan-21, Yoshida-21, Hyart-22, Alsallom-21, OrtegaTaberner-22}, they focused only on the topological characterization in terms of a line gap.
For example, the fate of interacting Hermitian topological phases to non-Hermitian perturbations was studied, such as non-Hermitian extensions of the fractional quantum Hall insulator~\cite{Yoshida-19, *Yoshida-20}, toric code~\cite{Guo-20, Matsumoto-21, Shackleton-20}, and topological Mott insulator~\cite{Zhang-20Mott, Liu-20, Xu-20}.
Thus, there remains a need for developing a theory of the intrinsic non-Hermitian topology in the many-body regime.
Furthermore, many-body topology of non-Hermitian systems is relevant to the topological characterization of Liouvillians appearing in master equations~\cite{Diehl-11, *Bardyn-13, Budich-15, *Bardyn-18, Gong-17, Lieu-20, Tonielli-20, Altland-21}.

In the Hermitian case, an important effect of many-body interactions is the reduction of topological phases in noninteracting systems.
For example, while one-dimensional Hermitian systems with chiral symmetry are characterized by the $\mathbb{Z}$ topological invariant in the single-particle regime, many-body interactions can reduce the band topology and replace the $\mathbb{Z}$ topological invariant with the $\mathbb{Z}_4$ one~\cite{Fidkowski-Kitaev-10, *Fidkowski-Kitaev-11}.
However, it has been unclear how many-body interactions affect the band topology of non-Hermitian systems.
In a recent work~\cite{Yoshida-21}, point-gap topology of zero-dimensional non-Hermitian systems protected by chiral symmetry was shown to be subject to reduction due to many-body interactions.
Meanwhile, it has remained unclear whether the intrinsic non-Hermitian topology in one dimension, which is relevant to exceptional points and the skin effect, reduces or survives in the presence of many-body interactions.

In this work, we develop a topological theory of non-Hermitian many-body systems.
In Sec.~\ref{sec: invariant}, we formulate a many-body topological invariant for non-Hermitian interacting systems in one dimension.
The many-body topological invariant is defined by the winding of the complex-valued many-body spectrum in terms of a $\mathrm{U} \left( 1 \right)$ gauge field (i.e., magnetic flux) and describes the nonequilibrium dynamics generated by the non-Hermitian operators.
We demonstrate that it is free from reduction in the presence of many-body interactions.
In Sec.~\ref{sec: band topology}, we discuss the relationship between the many-body topology and band topology of non-Hermitian systems.
We argue that Hermitization, which plays a key role in the characterization of non-Hermitian band topology, is no longer applicable in many-body systems.
In Sec.~\ref{sec: Hatano-Nelson}, we investigate the interacting Hatano-Nelson model as an illustrative example of the many-body topological invariant.
We find the unique complex-spectral winding and the concomitant skin effect induced by the interplay of non-Hermiticity and many-body interactions.
In Sec.~\ref{sec: conclusion}, we conclude this work and discuss several outlooks.

%%%%%%%%%%
\section{Many-body topological invariant}
    \label{sec: invariant}

We introduce a topological invariant $W = W \left( E \right)$ for a generic non-Hermitian operator $\hat{H}$ in one dimension and reference energy $E \in \mathbb{C}$.
Here, $\hat{H}$ can be either a bosonic or fermionic operator.
Suppose that $\hat{H}$ respects $\mathrm{U} \left( 1 \right)$ symmetry, i.e., $\hat{H}$ commutes with the total particle number operator:
\begin{equation}
    [ \hat{H}, \hat{N} ] = 0.
        \label{eq: N-commutation}
\end{equation}
As a result of $\mathrm{U} \left( 1 \right)$ symmetry, $\hat{H}$ can be block diagonalized according to the eigenvalue of $\hat{N}$.
Then, we consider the $N$-particle operator
\begin{equation}
    \hat{H}_{N} \coloneqq \hat{\cal P}_N \hat{H} \hat{\cal P}_N,
\end{equation}
where $\hat{\cal P}_N$ is the projector onto the subspace with the fixed particle number $N$.
To define a topological invariant, we need an energy gap.
Here, we assume
\begin{equation}
    \det\,[ \hat{H}_{N} - E ] \neq 0
\end{equation}
as a gap condition for $\hat{H}_{N}$.
This is a many-body generalization of the point gap in band theory~\cite{Gong-18, KSUS-19}.
Similarly to the noninteracting case, $\hat{H}_{N}$ with a point gap can be flattened to a unitary operator in the $N$-particle subspace.

Owing to $\mathrm{U} \left( 1 \right)$ symmetry, we can introduce a $\mathrm{U} \left( 1 \right)$ gauge field $A_{n, n+1}$ on the link between sites $n$ and $n+1$ even for the non-Hermitian operator.
From this local gauge field, a magnetic flux $\phi$ is given as $\phi \coloneqq \sum_{n=1}^{L} A_{n, n+1}$.
We further assume the presence of the point gap for arbitrary $\phi$, i.e.,
\begin{equation}
    \forall \phi \in \left[ 0, 2\pi \right)\quad\det\,[ \hat{H}_{N} \left( \phi \right) - E ] \neq 0.
\end{equation}
The complex spectrum of $\hat{H}_{N} \left( \phi \right)$ is independent of the gauge choice.
Furthermore, although the complex spectrum of $\hat{H}_{N} \left( \phi \right)$ generally depends on $\phi$, it is invariant in the presence of a unit flux $\phi = 2\pi$.
Consequently, the winding number of the determinant of $\hat{H}_{N} \left( \phi \right) - E$ in the complex plane is well defined under the adiabatic insertion of a unit magnetic flux.
Then, we can identify this complex-spectral winding number as a topological invariant $W = W \left( E \right)$.
More precisely, $W \left( E \right)$ is given as
\begin{equation}
    W \left( E \right) \coloneqq \oint_0^{2\pi} \frac{d\phi}{2\pi\ii} \frac{d}{d\phi} \log \det\,[ \hat{H}_N \left( \phi \right) - E ].
        \label{eq: W}
\end{equation}
Notably, $W \left( E \right)$ depends on the reference energy $E$, as well as the non-Hermitian operator $\hat{H}$.

This topological invariant is well defined even in the presence of many-body interactions and disorder.
In noninteracting systems (i.e., $N=1$), the topological invariant $W$ in Eq.~(\ref{eq: W}) was applied to non-Hermitian disordered systems that exhibit localization transitions~\cite{Gong-18, Longhi-19, Claes-21}.
We demonstrate that a similar complex-spectral winding number $W$ is well defined also for a non-Hermitian many-body operator.

Importantly, the topological invariant $W \left( E \right)$ is defined only by the complex spectrum.
In fact, $W \left( E \right)$ in Eq.~(\ref{eq: W}) can be written as
\begin{align}
    W \left( E \right) \coloneqq \sum_{n=1}^{D_N} \oint_0^{2\pi} \frac{d\phi}{2\pi\ii} \frac{d}{d\phi} \log\,[ E_{N}^{(n)} \left( \phi \right) - E ],
\end{align}
where $D_N$ is the dimension of the $N$-particle Hilbert space, and $E_{N}^{(n)} \left( \phi \right)$ is the $n$\,th eigenenergy of $\hat{H}_{N} \left( \phi \right)$.
This is contrasted with the topological invariants of Hermitian operators, which are formulated solely by their eigenstates.
For example, the many-body Chern number is defined by the ground-state wave function in the presence of a gauge flux~\cite{Niu-Thouless-Wu-85}.
While such state-based topological invariants can be defined also for non-Hermitian operators,
the spectral formulation of topological phases is a new type of topological characterization intrinsic to non-Hermitian operators.
While no topological phases appear in one-dimensional Hermitian systems without symmetry~\cite{HK-review, QZ-review, CTSR-review}, a topological invariant can be assigned to one-dimensional non-Hermitian systems without symmetry, as discussed above.
Topological invariants of Hermitian systems describe the static order of eigenstates including the ground states.
By contrast, topological invariants intrinsic to non-Hermitian systems describe the nonequilibrium dynamics described by non-Hermitian operators.
As discussed in Ref.~\cite{KSR-21}, the intrinsic non-Hermitian topology also requires a new formulation of topological field theory.

The topological invariant can vanish in the presence of certain symmetry.
For example, it vanishes when the non-Hermitian operator respects reciprocity
\begin{equation}
    \hat{\cal T} \hat{H}_{N}^{T} \left( \phi \right) \hat{\cal T}^{-1} = \hat{H}_{N} \left( - \phi \right),
\end{equation}
where $\hat{\cal T}$ is a unitary operator. 
In fact, in the presence of reciprocity, $W \left( E \right)$ in Eq.~(\ref{eq: W}) satisfies
\begin{align}
    W \left( E \right)
    &= \oint_0^{2\pi} \frac{d\phi}{2\pi\ii} \frac{d}{d\phi} \log \det\,[ \hat{\cal T} \hat{H}_N^{T} \left( -\phi \right) \hat{\cal T}^{-1} - E ] \nonumber \\
    &= \oint_{0}^{2\pi} \frac{d\phi}{2\pi\ii} \frac{d}{d\phi} \log \det\,[ \hat{H}_N \left( -\phi \right) - E ] \nonumber \\
    &= - W \left( E \right),
\end{align}
leading to $W \left( E \right) = 0$.
The vanishing winding number due to reciprocity is similar to the noninteracting regime~\cite{KSUS-19, KOS-20}.

Before proceeding, we provide several further remarks.
First, the topological invariant is well defined for a generic non-Hermitian many-body operator $\hat{H}$. 
For example, $\hat{H}$ can be a non-Hermitian Hamiltonian that effectively describes open quantum systems subject to postselection~\cite{Daley-review, Ashida-review}. 
In addition, $\hat{H}$ can be a Liouvillian of a master equation~\cite{Breuer-textbook, Rivas-textbook}.
    
Second, $\mathrm{U} \left( 1 \right)$ symmetry in  Eq.~(\ref{eq: N-commutation}) enables the introduction of the $\mathrm{U} \left( 1 \right)$ gauge field.
In Hermitian systems, $\mathrm{U} \left( 1 \right)$ symmetry also leads to conservation of the particle number.
In non-Hermitian systems, by contrast, this is not necessarily the case.
In the Heisenberg picture, the particle number operator $\hat{N}$ evolves as
\begin{equation}
\hat{N} \left( t \right) = e^{\ii \hat{H}^{\dag} t} \hat{N} e^{-\ii \hat{H} t}
\end{equation}
under the non-Hermitian Hamiltonian $\hat{H}$.
Thus, we have
\begin{equation}
\ii \frac{d\hat{N}}{dt} = \hat{N} \hat{H} - \hat{H}^{\dag} \hat{N},  
\end{equation}
and the conservation of the particle number (i.e., $d\hat{N}/dt = 0$) is given by
\begin{equation}
\hat{N} \hat{H} - \hat{H}^{\dag} \hat{N} = 0.
        \label{eq: N-conservation}
\end{equation}
While Eqs.~(\ref{eq: N-commutation}) and (\ref{eq: N-conservation}) are equivalent to each other for $\hat{H} = \hat{H}^{\dag}$, this is not the case for $\hat{H} \neq \hat{H}^{\dag}$.
    
Next, we emphasize the importance of the block diagonalization according to $\hat{N}$.
In the presence of unitary symmetry that commutes with $\hat{H}$ [i.e., Eq.~(\ref{eq: N-commutation})], each block with fixed $N$ is independent and cannot interact with each other.
Consequently, we have to define the topological invariant for each subspace with fixed $N$.
If we considered the complex-spectral winding without block diagonalization, it would be meaningless and irrelevant to the skin effect.
Even in band theory, the block diagonalization is needed to understand the skin effect of non-Hermitian systems with commutative unitary symmetry~\cite{Okuma-19, OKSS-20, KOS-20}.
    
Finally, when the dimension of the Hilbert space and the degree of the point gap 
(i.e., $\min_{\phi\in\left[0,2\pi\right)} | \det\,[ \hat{H}_{N} \left( \phi \right) - E ] |$)
are sufficiently large, the integrand $\partial_{\phi} \log \det\,[ \hat{H}_{N} \left( \phi \right) - E ]$ in Eq.~(\ref{eq: W}) is expected to be independent of $\phi$.
Then, the topological invariant $W$ in Eq.~(\ref{eq: W}) is simplified to
    \begin{align}
    W \left( E \right) 
    &\simeq - \ii \frac{d}{d\phi} \log \det\,[ \hat{H}_N \left( \phi \right) - E ],
        \label{eq: W - simplified}
\end{align}
where the $\phi$ derivative can be taken for arbitrary $\phi$.
This simplification is similar to the Niu-Thouless-Wu formula for the many-body Chern number~\cite{Niu-Thouless-Wu-85, Kudo-19}.
In Sec.~\ref{sec: Hatano-Nelson}, we confirm this simplification for the interacting Hatano-Nelson model.

%%%%%%%%%%
\section{Relationship with band topology}
    \label{sec: band topology}

The intrinsic non-Hermitian topological phases were generally formulated for noninteracting systems in terms of band theory~\cite{Gong-18, KSUS-19, OKSS-20}.
The many-body topological invariant $W$ in Eq.~(\ref{eq: W}) reduces to the band topology for a single particle $N=1$ and in the presence of translation invariance.
The single-particle Hamiltonian $\hat{H}_1$ with translation invariance is diagonalized in momentum space as
\begin{align}
    \hat{H}_1 = \sum_{k \in  \mathrm{BZ}} \hat{c}_{k}^{\dag} H \left( k \right) \hat{c}_{k},
\end{align}
with the one-dimensional Brillouin zone
\begin{equation}
    \mathrm{BZ} \coloneqq \left\{ 0, \frac{2\pi}{L}, \frac{4\pi}{L}, \cdots, \frac{2\pi \left( L-1 \right)}{L} \right\} 
\end{equation}
and the Bloch Hamiltonian $H \left( k \right)$.
In the presence of a magnetic flux $\phi$, the Hamiltonian reads
\begin{align}
    \hat{H}_1 \left( \phi \right) = \sum_{k \in  \mathrm{BZ}} \hat{c}_{k-\phi/L}^{\dag} H \left( k-\phi/L \right) \hat{c}_{k-\phi/L},
\end{align}
where the gauge field is chosen to be uniform (i.e., $A_{n, n+1} \coloneqq \phi/L$).
Then, we have
\begin{equation}
    \det\,[ \hat{H}_1 \left( \phi \right) - E ]
    = \prod_{k \in  \mathrm{BZ}} \det\,[ H \left( k - \phi/L \right) - E ],
\end{equation}
and hence the topological invariant $W \left( E \right)$ in Eq.~(\ref{eq: W}) reduces to 
\begin{align}
    W \left( E \right)
    &= \sum_{k' \in  \mathrm{BZ}} \oint_{0}^{2\pi} \frac{d\phi}{2\pi\ii} \frac{d}{d\phi} \log \det\,[ H \left( k' - \phi/L \right) - E ] \nonumber \\
    &= - \sum_{k' \in  \mathrm{BZ}} \oint_{k'-2\pi/L}^{k'} \frac{dk}{2\pi\ii} \frac{d}{dk} \log \det\,[ H \left( k \right) - E ]\quad \nonumber \\
    &= - \oint_{0}^{2\pi} \frac{dk}{2\pi\ii} \frac{d}{dk} \log \det\,[ H \left( k \right) - E ],
\end{align}
which reproduces the topological invariant in band theory~\cite{Gong-18, KSUS-19}.
We note that $k$ in the second equality is introduced by $k \coloneqq k' - \phi/L$ to replace the magnetic flux $\phi$ with the momentum $k$.

Notably, the band topology of noninteracting non-Hermitian Hamiltonians is understood by Hermitization~\cite{Feinberg-97, Roy-17, Gong-18, KSUS-19}.
For a given noninteracting non-Hermitian Hamiltonian $H$, a noninteracting Hermitian Hamiltonian $\tilde{H}$ in the doubled Hilbert space can be constructed as
\begin{equation}
    \tilde{H} \coloneqq \begin{pmatrix}
    0 & H \\
    H^{\dag} & 0
    \end{pmatrix},
\end{equation}
where the reference energy $E$ is set to zero.
By construction, the extended Hermitian Hamiltonian $\tilde{H}$ respects chiral symmetry
\begin{equation}
    \tau_z \tilde{H} \tau_z^{-1} = - \tilde{H}
\end{equation}
with a Pauli matrix $\tau_z$.
For example, when we Hermitize the Hatano-Nelson model $H$~\cite{Hatano-Nelson-96, *Hatano-Nelson-97}, we obtain the Su-Schrieffer-Heeger model $\tilde{H}$~\cite{SSH-79}; 
the complex-spectral winding number of $H$ coincides with the winding number of the eigenstates in $\tilde{H}$.
In this manner, non-Hermitian band topology in terms of a point gap can be generally classified on the basis of Hermitization~\cite{Gong-18, KSUS-19, ZL-19}.
Hermitization plays a key role also in the skin effect~\cite{OKSS-20} and Anderson localization~\cite{Luo-21} of noninteracting non-Hermitian systems.

Now, suppose that Hermitization were valid in the same manner even in interacting systems.
Then, the many-body topology of one-dimensional non-Hermitian systems would be associated with the many-body topology of one-dimensional Hermitian systems with chiral symmetry.
Here, many-body interactions reduce the band topology of chiral-symmetric Hermitian systems characterized by the $\mathbb{Z}$ topological invariant to the $\mathbb{Z}_4$ topology~\cite{Fidkowski-Kitaev-10, *Fidkowski-Kitaev-11}.
Thus, the band topology of noninteracting non-Hermitian systems would also be reduced from $\mathbb{Z}$ to $\mathbb{Z}_4$ because of many-body interactions.
However, this would contradict the $\mathbb{Z}$ topological invariant well defined even for non-Hermitian many-body systems, as shown in Sec.~\ref{sec: invariant}.

This discussion implies that Hermitization is no longer valid in the many-body regime in the same manner as the noninteracting regime.
While Hermitization maps a noninteracting non-Hermitian Hamiltonian to a noninteracting Hermitian Hamiltonian in the doubled single-particle Hilbert space, it cannot preserve the structure of the many-body Hilbert space.
The correspondence between non-Hermitian systems and chiral-symmetric Hermitian systems is unique to the noninteracting regime and breaks down in the many-body regime.
Even if we may associate a non-Hermitian many-body system $\hat{H}$ with another Hermitian many-body system $\hat{\tilde H}$, chiral symmetry should not be respected by $\hat{\tilde H}$.

We note in passing that unitary operators can be mapped to ground states of Hermitian Hamiltonians in the double Hilbert space~\cite{Glorioso-19, *Liu-Ryu-20}.
In the single-particle regime, a non-Hermitian operator with a nontrivial topological invariant $W \neq 0$, including the Hatano-Nelson model~\cite{Hatano-Nelson-96, *Hatano-Nelson-97}, can be flattened to the unitary translation operator.
However, a single-particle unitary operator cannot be generalized to many-body operators in a unique manner, and an important difference arises between unitary operators and generic non-Hermitian operators in the many-body regime.
For unitary operators, the spectrum lies on the unit circle in the complex plane and retains a point gap around $E=0$ in all the $N$-particle subspaces;
for generic non-Hermitian operators, even if a point gap is open in a specific $N$-particle subspace, this point gap may be closed for another $N'$-particle subspace.
For example, in the Hatano-Nelson model, while a point gap is open around $E=0$ in the single-particle subspace, no point gap is open around $E = 0$ in the many-particle subspaces (see Sec.~\ref{sec: Hatano-Nelson} for details).
Consequently, generic non-Hermitian operators cannot be flattened to unitary operators in arbitrary $N$-particle subspaces.

The topological invariant $W$ in Eq.~(\ref{eq: W}) may be similar to the topological invariant of zero-dimensional Hermitian systems, which is defined by the $\mathrm{U} \left( 1 \right)$ charge (i.e., particle number) of the ground state, rather than one-dimensional Hermitian systems with chiral symmetry.
This is compatible with the topological field theory---(1+0)-dimensional Chern-Simons theory---for one-dimensional non-Hermitian systems~\cite{KSR-21}.

%%%%%%%%%%
\section{Example: interacting Hatano-Nelson model}
    \label{sec: Hatano-Nelson}
    
While the many-body topological invariant $W$ in Eq.~(\ref{eq: W}) reduces to the band topology for noninteracting Hamiltonians with translation invariance, its relevance is unclear for many-particle cases $N \geq 2$.
To understand the role of $W$ in non-Hermitian many-body systems, we investigate the interacting Hatano-Nelson model
\begin{align}
    \hat{H} 
    &= \sum_{n=1}^{L} \left( - \frac{1+\gamma}{2} \hat{c}_{n+1}^{\dag} \hat{c}_{n} - \frac{1-\gamma}{2} \hat{c}_{n}^{\dag} \hat{c}_{n+1} \right. \nonumber \\
    &\qquad\qquad\qquad\qquad\qquad + \left. U \hat{c}_{n}^{\dag} \hat{c}_{n} \hat{c}_{n+1}^{\dag} \hat{c}_{n+1} \right).
\end{align}
Here, $\hat{c}_{n}$ ($\hat{c}_{n}^{\dag}$) is an annihilation (creation) operator of a spinless fermion at site $n$.
In addition, $\gamma \in \mathbb{R}$ is the degree of non-Hermiticity, and $U \in \mathbb{R}$ is the strength of the two-body interaction.
In the absence of the interaction (i.e., $U = 0$), the model reduces to the Hatano-Nelson model~\cite{Hatano-Nelson-96, *Hatano-Nelson-97}, which is a prototypical model that exhibits intrinsic non-Hermitian topology in band theory~\cite{Gong-18, KSUS-19}.
Similar fermionic interacting models were also investigated in Refs.~\cite{Hamazaki-19, Mu-20, Orito-21}.
Furthermore, similar spin models (i.e., XXZ chains with asymmetric XX coupling) were investigated in Refs.~\cite{Albertini-96, Fukui-98Nucl, Nakamura-06}. 
Below, we study the topological phase and skin effect of the interacting Hatano-Nelson model.
In the presence of a magnetic flux $\phi \in \left[ 0, 2\pi \right)$, the Hamiltonian reads
\begin{align}
    \hat{H} 
    &= \sum_{n=1}^{L} \left( - e^{\ii \phi/L} \frac{1+\gamma}{2} \hat{c}_{n+1}^{\dag} \hat{c}_{n} - e^{-\ii \phi/L} \frac{1-\gamma}{2} \hat{c}_{n}^{\dag} \hat{c}_{n+1} \right. \nonumber \\
    &\qquad\qquad\qquad\qquad\qquad\quad + \left. U \hat{c}_{n}^{\dag} \hat{c}_{n} \hat{c}_{n+1}^{\dag} \hat{c}_{n+1} \right),
\end{align}
where the gauge field is chosen to be uniform (i.e., $A_{n, n+1} \coloneqq \phi/L$).

%%%%%
\subsection{$N=1$ (single particle)}

\begin{figure}[t]
\centering
\includegraphics[width=86mm]{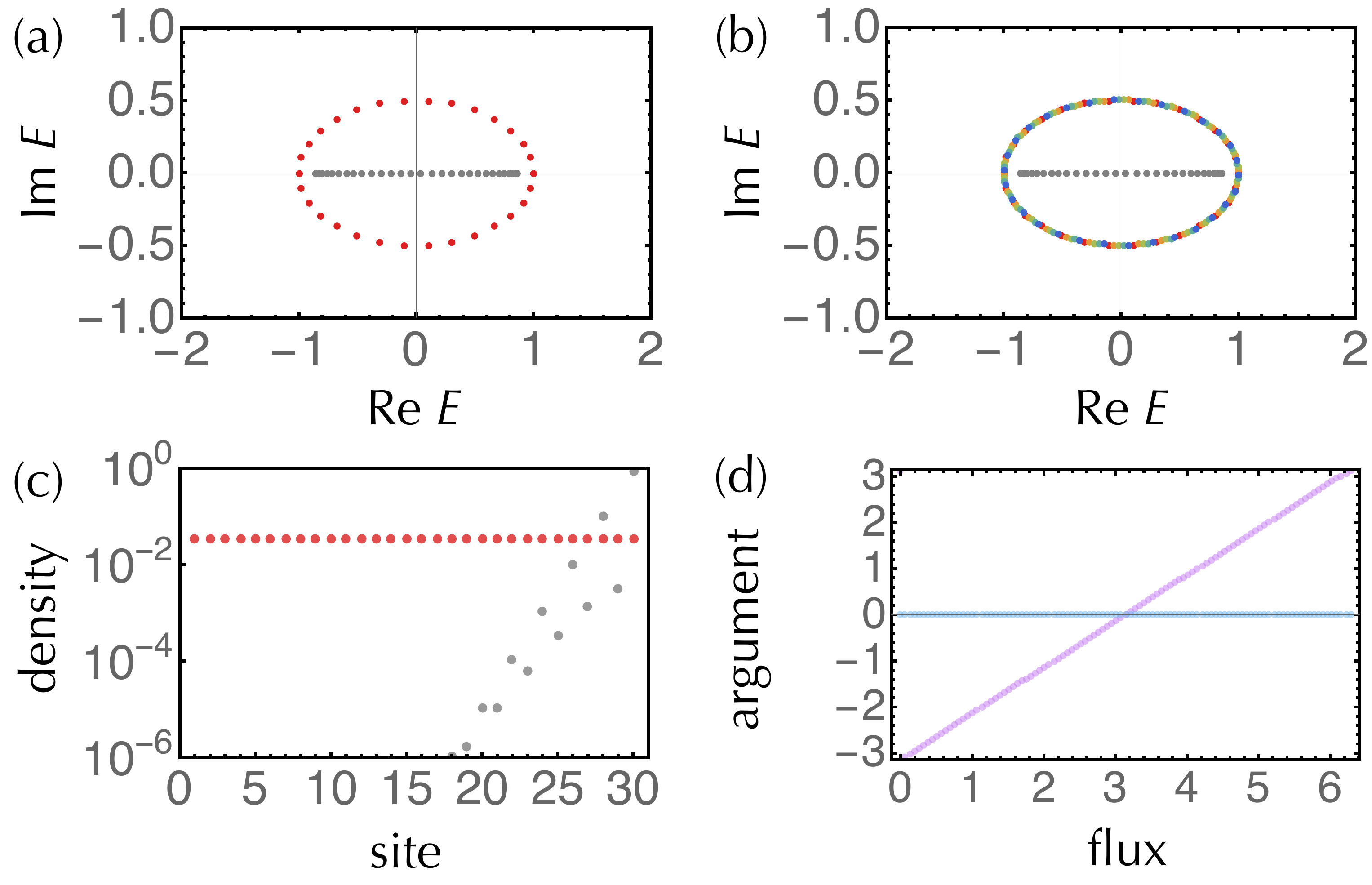} 
\caption{Interacting Hatano-Nelson model with one particle ($L=30$, $N=1$, $\gamma=0.5$). (a)~Complex spectra under the periodic boundary conditions (red dots) and open boundary conditions (black dots). (b)~Complex spectra under the periodic and open boundary conditions in the presence of the flux $\phi \in \{ 0, 2\pi/5, 4\pi/5, 6\pi/5, 8\pi/5 \}$. (c)~Spatial distributions of particle numbers for an eigenstate with $E = -0.10 - 0.50\ii$ under the periodic boundary conditions (red dots) and an eigenstate with $E=-0.04$ under the open boundary conditions (black dots). (d)~Arguments of the determinants of $\hat{H}_{1} \left( \phi \right) - E$ as a function of the flux $\phi$ for $E = 0$ (purple dots, $W = 1$) and $E = -1.5$ (blue dots, $W = 0$).}
	\label{fig: N1}
\end{figure}

For clarity, we begin with the single-particle case $N=1$.
In the single-particle sector, the two-body interaction $U \hat{c}_{n}^{\dag} \hat{c}_{n} \hat{c}_{n+1}^{\dag} \hat{c}_{n+1}$ is irrelevant, and the Hamiltonian with periodic boundaries reads in momentum space
\begin{align}
    \hat{H}_{1} = \sum_{k \in {\rm BZ}} E \left( k \right) \hat{c}_{k}^{\dag} \hat{c}_{k}
\end{align}
with the complex-energy dispersion
\begin{align}
    E \left( k \right) 
    &\coloneqq - \frac{1+\gamma}{2} e^{-\ii k} - \frac{1-\gamma}{2} e^{\ii k} \nonumber \\
    &= - \cos k + \ii \gamma \sin k.
        \label{eq: energy dispersion}
\end{align}
The spectrum forms a loop in the complex-energy plane [Fig.~\ref{fig: N1}\,(a, b)].
Because of this loop structure of the complex spectrum, we have $W_1 \left( E \right) = \mathrm{sgn} \left( \gamma \right)$ [$W_1 \left( E \right) = 0$] when the reference energy $E$ is inside (outside) the loop [Fig.~\ref{fig: N1}\,(d)].
As demonstrated in Refs.~\cite{Zhang-20, OKSS-20}, the complex-spectral winding number $W$ leads to the skin effect under the open boundary conditions.
Consistently, the open-boundary spectrum is drastically different from the periodic-boundary spectrum, and the eigenstates are localized at the edge [Fig.~\ref{fig: N1}\,(c)].
It is notable that $-\ii \partial_{\phi} \log \det\,[ \hat{H}_1 \left( \phi \right) - E ] = \partial_{\phi}\,\mathrm{arg}\,\det\,[ \hat{H}_1 \left( \phi \right) - E ]$ is almost independent of $\phi$ when the degree of the point gap is sufficiently large.
The simplified formula in Eq.~(\ref{eq: W - simplified}) is thus valid even in the single-particle case.

%%%%%
\subsection{$N=2$}

\begin{figure}[t]
\centering
\includegraphics[width=86mm]{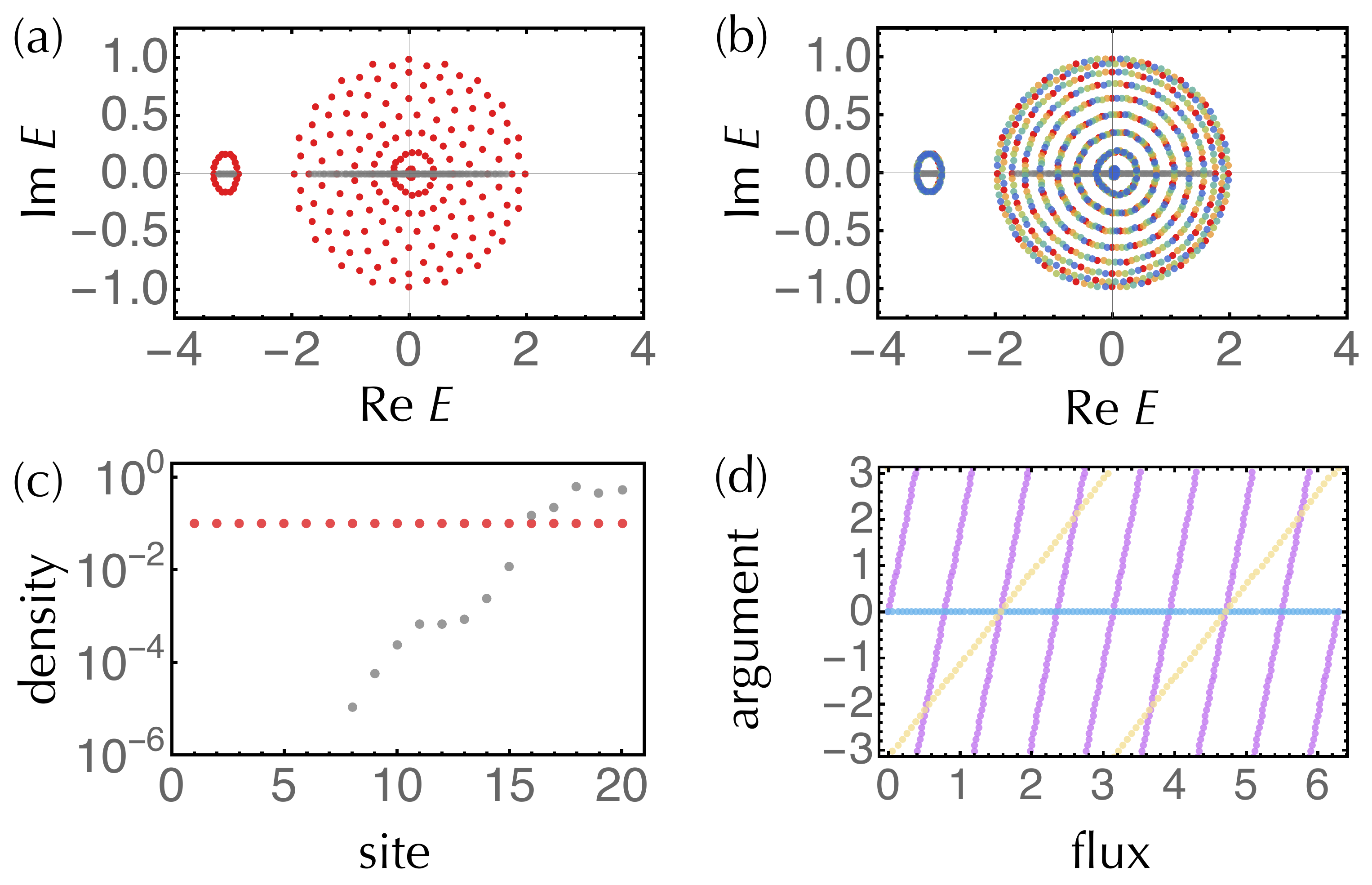} 
\caption{Interacting Hatano-Nelson model with two particles ($L=20$, $N=2$, $\gamma=0.5$, $U=-3.0$). (a)~Complex spectra under the periodic boundary conditions (red dots) and open boundary conditions (black dots). (b)~Complex spectra under the periodic and open boundary conditions in the presence of the flux $\phi \in \{ 0, 2\pi/5, 4\pi/5, 6\pi/5, 8\pi/5 \}$. (c)~Spatial distributions of particle numbers for an eigenstate with $E = -0.02$ under the periodic boundary conditions (red dots) and an eigenstate with $E = -0.05$ under the open boundary conditions (black dots). (d)~Arguments of the determinants of $\hat{H}_{2} \left( \phi \right) - E$ as a function of the flux $\phi$ for $E = -0.2$ (purple dots, $W = 8$), $E=-2.5$ (blue dots, $W=0$), and $E = -3.2$ (yellow dots, $W = 2$).}
	\label{fig: N2}
\end{figure}

In the two-particle case $N=2$, we numerically diagonalize the non-Hermitian Hamiltonian $\hat{H}_2 \left( \phi \right)$ and obtain the topological invariant $W \left (E \right)$ (Fig.~\ref{fig: N2}).
In these numerical calculations, the attractive interaction $U < 0$ is considered.
First, the complex spectrum includes multiple layers of loops around the origin [Fig.~\ref{fig: N2}\,(a, b)].
Consequently, we can have a large winding number $W$ [Fig.~\ref{fig: N2}\,(d)], which is unfeasible in the single-particle case.
This multilayer structure is typical behavior of many-particle non-Hermitian Hamiltonians and does not actually require the many-body interaction. 
In the absence of the interaction (i.e., $U = 0$), the complex spectrum of the two-particle Hamiltonian $\hat{H}_2$ is given as
\begin{equation}
    E \left( k \right) + E \left( k' \right),
\end{equation}
where $k$ and $k'$ are independent momenta, and $E \left( k \right)$ is the single-particle energy dispersion in Eq.~(\ref{eq: energy dispersion}).
As a result, the complex spectrum forms the following disk for $L \to \infty$:
\begin{equation}
    \left( \frac{\mathrm{Re}\,E}{2}\right)^2 + \left( \frac{\mathrm{Im}\,E}{2\gamma} \right)^2 \leq 1.
\end{equation}
The multilayer structure numerically obtained for an interacting and finite system is reminiscent of the disk for the noninteracting and infinite system.
It should be noted that the presence of the point gap between the loops is due to the finite-size effect.
As the system size $L$ increases, this point gap gets smaller.
Meanwhile, the number of the loops, as well as the complex-spectral winding numbers, gets larger.
For the infinite-size limit $L\to \infty$, the point gap vanishes, and the winding numbers diverge and are no longer well defined.
While a point gap is open around $E=0$ even for $L \to \infty$ in the single-particle case $N=1$, no point gap is open around $E=0$ in the two-particle case $N=2$.

The skin effect occurs also in the two-particle case [Fig.~\ref{fig: N2}\,(c)].
The two-body interaction complicates the localization of skin modes.
Still, skin modes appear only for the energy with $W \left( E \right) \neq 0$.
The correspondence between the topological invariant $W$ and the skin effect is hitherto proved only for the single-particle case~\cite{Zhang-20, OKSS-20}.
Our numerical calculations may suggest a similar relationship even in non-Hermitian many-body systems.

Another remarkable feature in the two-particle spectrum is the appearance of an additional loop isolated from the multiple loops around the origin in the complex-energy plane [Fig.~\ref{fig: N2}\,(a, b)].
Such an isolated loop appears only for a sufficiently large interaction $U$ and is characterized by the winding number $W = 2\,\mathrm{sgn} \left( \gamma \right)$, and consequently, the skin effect occurs under the open boundary conditions.
In contrast to the multilayer structure around the origin, the point gap of this isolated loop is open even for $L \to \infty$.
As discussed below, this isolated loop can be understood by a second-order perturbation theory in terms of $1/U$. 
For a sufficiently strong many-body interaction, the energy separation occurs in the many-body spectrum.
In the additional presence of non-Hermiticity $\gamma$, a point gap is open for this separated many-body spectrum, which leads to the formation of the isolated loop. 
Thus, the isolated loop in the two-particle complex spectrum originates from the interplay between many-body interactions and non-Hermiticity.
We note in passing that it may be related to the cluster structure in the complex spectrum of random Liouvillians~\cite{Wang-Luitz-20};
it is interesting to investigate the spectral structure of non-Hermitian many-body systems in the presence of more general many-body interactions.

Now, we derive the effective Hamiltonian $\hat{H}_{\rm eff}$ when the interaction term 
\begin{align}
    \hat{H}_{\rm int} 
    &\coloneqq U \sum_{n=1}^{L} \hat{c}_{n}^{\dag} \hat{c}_{n} \hat{c}_{n+1}^{\dag} \hat{c}_{n+1}
\end{align}
is much larger than the hopping term 
\begin{align}
    \hat{H}_{\rm hop} 
    &\coloneqq \sum_{n=1}^{L} \left( - \frac{1+\gamma}{2} \hat{c}_{n+1}^{\dag} \hat{c}_{n} - \frac{1-\gamma}{2} \hat{c}_{n}^{\dag} \hat{c}_{n+1} \right).
\end{align}
We obtain $\hat{H}_{\rm eff}$ by a perturbation theory for $\left| U \right| \gg 1$.
For the two-particle sector $N=2$, the spectrum of $\hat{H}_{\rm int}$ consists of $E = 0$ and $E = U$.
Here, we focus on $E = U$ to account for the separate loop induced by the many-body interaction.
For $E = U$, we have $L$ eigenstates 
\begin{equation}
    | n \rangle\!\rangle \coloneqq \hat{c}_{n}^{\dag} \hat{c}_{n+1}^{\dag} \ket{\rm vac}\quad \left( n = 1,2, \cdots L \right)
\end{equation}
with the fermionic vacuum $\ket{\rm vac}$.
Then, in the presence of the hopping term $\hat{H}_{\rm hop}$, the effective Hamiltonian $\hat{H}_{\rm eff}$ is perturbatively obtained as
\begin{align}
    &\hat{H}_{\rm eff}
    = E + \hat{\cal P}_{\rm int} \hat{H}_{\rm hop} \hat{\cal P}_{\rm int} \nonumber \\
    &\quad+ \hat{\cal P}_{\rm int} \hat{H}_{\rm hop}\,( E - \hat{H}_{\rm int} )^{-1} \hat{H}_{\rm hop} \hat{\cal P}_{\rm int}
    + \mathcal{O}\,( \hat{H}_{\rm hop}^3 ),
        \label{eq: perturbation formula}
\end{align}
where $\hat{\cal P}_{\rm int} \coloneqq \sum_{n=1}^{L} | n \rangle\!\rangle \langle\!\langle n |$ is the projector onto the eigenspace of $\hat{H}_{\rm int}$ [see, for example, Sec.~10.1 of Ref.~\cite{Tasaki-textbook} for a derivation of Eq.~(\ref{eq: perturbation formula})].
The first-order contribution vanishes, i.e., $\hat{\cal P}_{\rm int} \hat{H}_{\rm hop} \hat{\cal P}_{\rm int} = 0$.
On the other hand, the second-order contribution is computed explicitly as
\begin{align}
    &\langle\!\langle m | \hat{H}_{\rm hop}\,( E - \hat{H}_{\rm int} )^{-1} \hat{H}_{\rm hop} | n \rangle\!\rangle \nonumber \\
    &\quad = \frac{1-\gamma^2}{2U} \delta_{m, n}
    + \frac{\left( 1+\gamma \right)^2}{4U} \delta_{m, n+1}
    + \frac{\left( 1-\gamma \right)^2}{4U} \delta_{m, n-1}.
\end{align}
Thus, the effective Hamiltonian $\hat{H}_{\rm eff}$ in Eq.~(\ref{eq: perturbation formula}) is 
\begin{align}
    &\hat{H}_{\rm eff} 
    \simeq U + \frac{1-\gamma^2}{2U} \nonumber \\
    &\quad + \sum_{n=1}^{L} \left[ \frac{\left( 1+\gamma \right)^2}{4U} | n+1 \rangle\!\rangle \langle\!\langle n |
    + \frac{\left( 1-\gamma \right)^2}{4U} | n \rangle\!\rangle \langle\!\langle n+1 | \right].
\end{align}

The obtained effective Hamiltonian $\hat{H}_{\rm eff}$ is similar to the single-particle Hatano-Nelson model whose hopping amplitude from the left to the right (from the right to the left) is $\left( 1+ \gamma \right)^2/4U$ [$\left( 1- \gamma \right)^2/4U$].
Hence, the spectrum is given as
\begin{align}
    E &\simeq U + \frac{1-\gamma^2}{2U}+ \frac{\left( 1+\gamma \right)^2}{4U} e^{-\ii \theta} + \frac{\left( 1-\gamma \right)^2}{4U} e^{\ii \theta} \nonumber \\
    &= U + \frac{1-\gamma^2}{2U}+ \frac{1+\gamma^2}{2U} \cos \theta - \ii \frac{\gamma}{U} \sin \theta
        \label{eq: H2 - PBC - spectrum}
\end{align}
with $\theta \in \left[ 0, 2\pi \right)$.
This is consistent with the numerical results in Fig.~\ref{fig: N2}.
In the presence of the gauge flux $\phi$, the effective Hamiltonian reads
\begin{align}
    &\hat{H}_{\rm eff} \left( \phi \right) 
    \simeq U + \frac{1-\gamma^2}{2U} \nonumber \\
    &\qquad\quad + \sum_{n=1}^{L} \left[ e^{-2\ii\phi/L} \frac{\left( 1+\gamma \right)^2}{4U} | n+1 \rangle\!\rangle \langle\!\langle n | \right. \nonumber \\
    &\qquad\qquad\qquad\qquad \left. + e^{2\ii\phi/L} \frac{\left( 1-\gamma \right)^2}{4U} | n \rangle\!\rangle \langle\!\langle n+1 | \right],
\end{align}
which leads to the winding number $W = 2\,\mathrm{sgn} \left( \gamma \right)$ for the reference energy inside the loop.
The doubled winding number compared to the single-particle case is a unique feature of the many-body topological invariant. 

Under the open boundary conditions, the effective Hamiltonian reads
\begin{align}
    &\hat{H}_{\rm eff} 
    \simeq U + \frac{1-\gamma^2}{2U} \left( 1 - \frac{| 1 \rangle\!\rangle \langle\!\langle 1 | + | L-1 \rangle\!\rangle \langle\!\langle L-1 |}{2}\right)  \nonumber \\
    &\quad + \sum_{n=1}^{L-2} \left[ \frac{\left( 1+\gamma \right)^2}{4U} | n+1 \rangle\!\rangle \langle\!\langle n |
    + \frac{\left( 1-\gamma \right)^2}{4U} | n \rangle\!\rangle \langle\!\langle n+1 | \right].
\end{align}
The spectrum of this effective Hamiltonian is obtained as
\begin{equation}
    E = U + \frac{1-\gamma^2}{2U} \left( 1 + \cos \theta \right)
\end{equation}
with $\theta = m\pi/\left( L-1 \right)$ ($m = 1, 2, \cdots, L-1$).
This is clearly different from the spectrum for the periodic boundary conditions [i.e., Eq.~(\ref{eq: H2 - PBC - spectrum})], which is a signature of the skin effect.
In fact, the corresponding right eigenstate is
\begin{equation}
    \propto \sum_{n=1}^{L-1} \left( \frac{1+\gamma}{1-\gamma} \right)^{n} \sin \left[ \left( n-\frac{1}{2} \right) \theta \right] | n \rangle\!\rangle,
\end{equation}
localized at the right (left) edge for $\gamma > 0$ ($\gamma < 0$).
Similarly to the single-particle case, the skin effect under the open boundary conditions is consistent with the spectral winding number $W = 2\,\mathrm{sgn} \left( \gamma \right)$ under the periodic boundary conditions.

%%%%%
\subsection{$N=3$}

\begin{figure}[t]
\centering
\includegraphics[width=86mm]{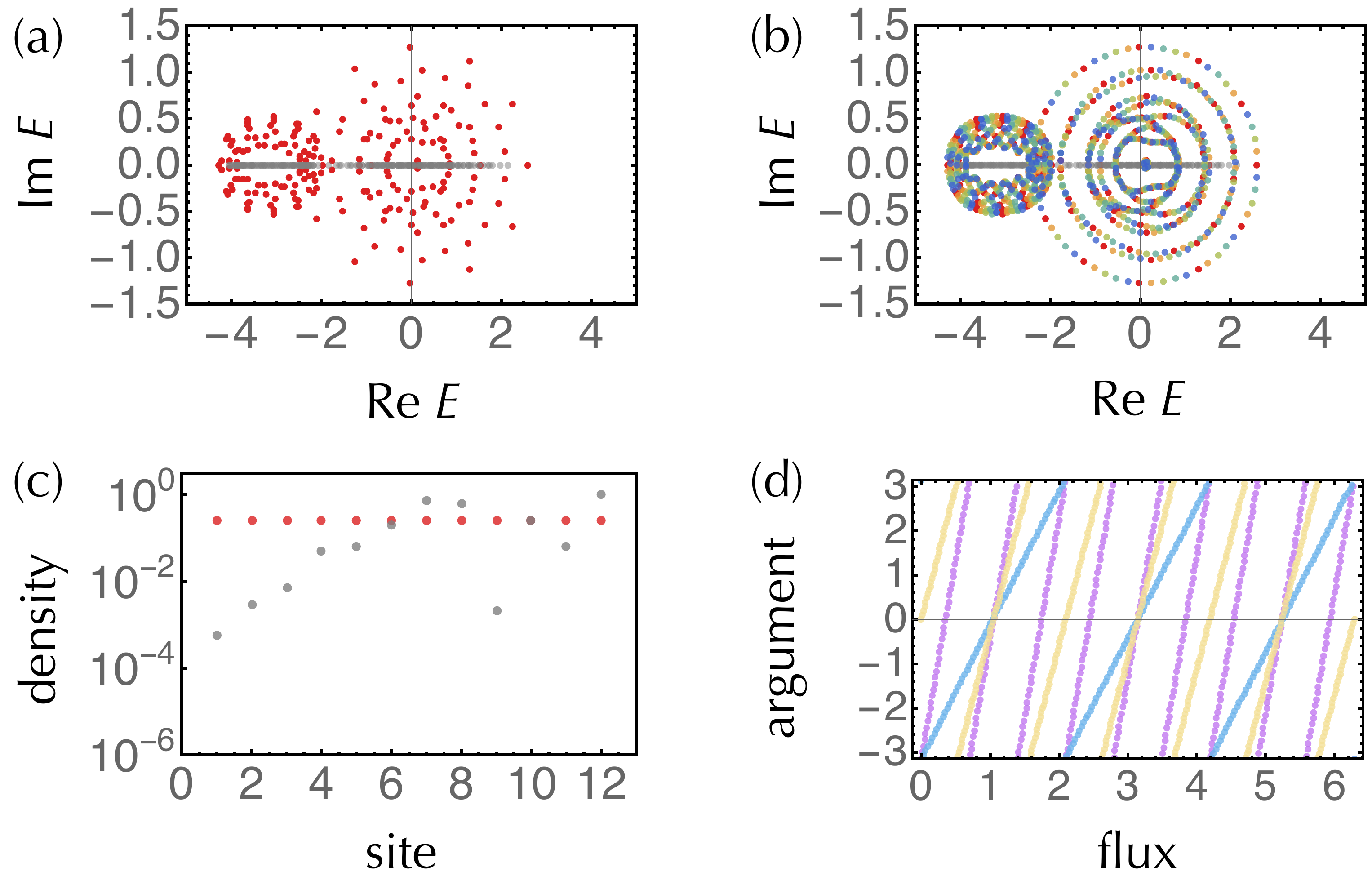} 
\caption{Interacting Hatano-Nelson model with three particles ($L=12$, $N=3$, $\gamma=0.5$, $U=-3.0$). (a)~Complex spectra under the periodic boundary conditions (red dots) and open boundary conditions (black dots). (b)~Complex spectra under the periodic and open boundary conditions in the presence of the flux $\phi \in \{ 0, 2\pi/5, 4\pi/5, 6\pi/5, 8\pi/5 \}$. (c)~Spatial distributions of particle numbers for an eigenstate with $E = -0.004 - 0.28\ii$ under the periodic boundary conditions (red dots) and an eigenstate with $E=-0.005$ under the open boundary conditions (black dots). (d)~Arguments of the determinants of $\hat{H}_{3} \left( \phi \right) - E$ as a function of the flux $\phi$ for $E = -0.2$ (purple dots, $W = 9$), $E=-1.6$ (blue dots, $W=3$), and $E = -3.0$ (yellow dots, $W = 6$).}
	\label{fig: N3}
\end{figure}

Figure~\ref{fig: N3} shows the numerically obtained complex spectra, eigenstates, and winding numbers for the three-particle case $N=3$. 
While the spectrum becomes more complicated, it is qualitatively similar to the two-particle spectrum:
multilayer loops around the origin and interaction-induced separate loops.
Similarly to the two-particle case, for the infinite-size limit $L\to\infty$, the point gap between the multilayer loops vanishes, and the concomitant winding numbers are ill defined.
Notably, the separate loops consist of multiple layers in contrast to the two-particle case.
Similarly to the fewer-particle case, the skin effect occurs also in the three-particle case.
The skin modes for the open boundary conditions seem to appear only in the energy regions characterized by the nonzero topological invariant for the periodic boundary conditions.
The simplified formula in Eq.~(\ref{eq: W - simplified}) is also valid.

%%%%%
\subsection{$N=5$ (half filling)}

\begin{figure}[t]
\centering
\includegraphics[width=86mm]{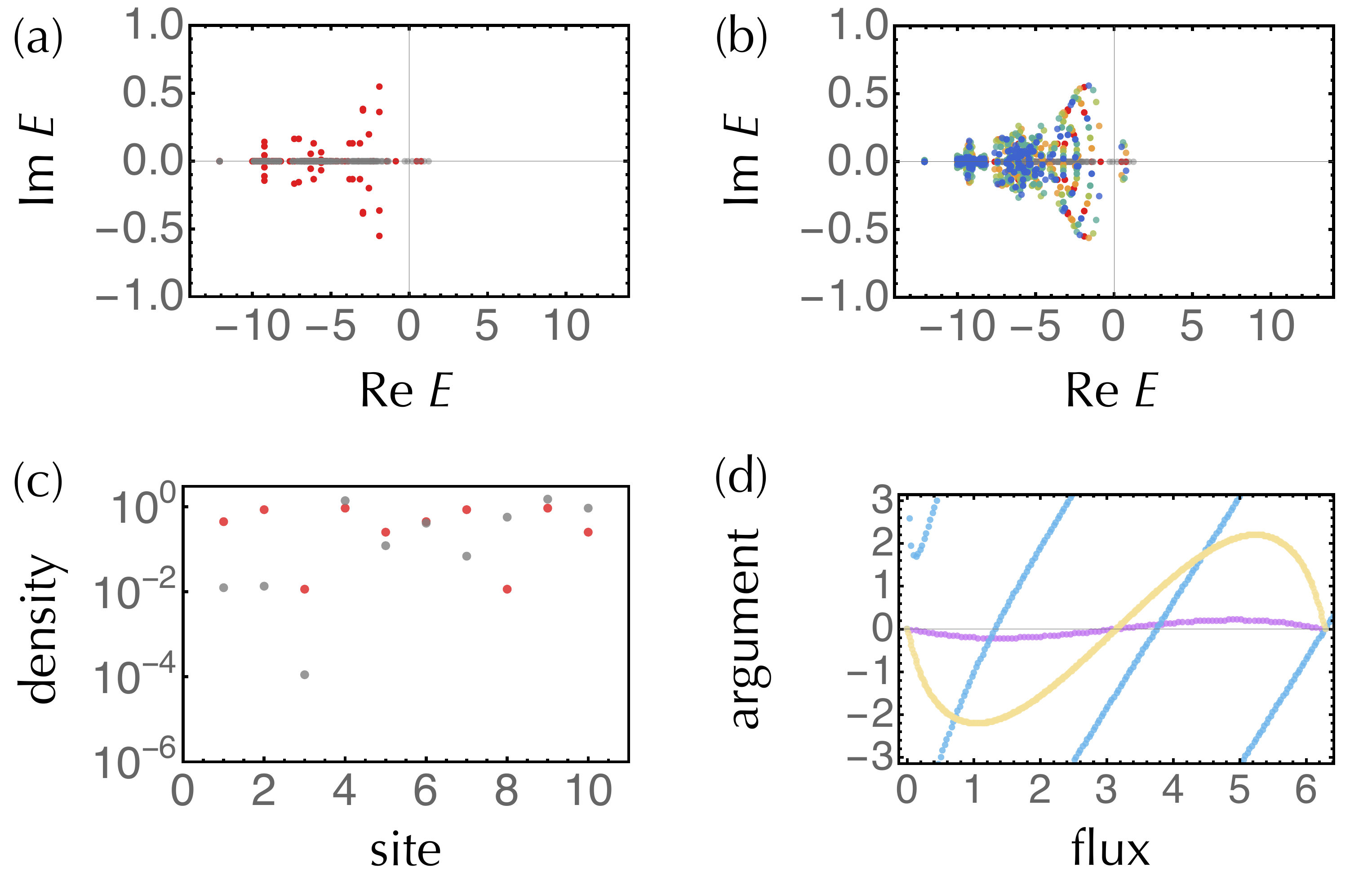} 
\caption{Interacting Hatano-Nelson model with five particles ($L=10$, $N=5$, $\gamma=0.5$, $U=-3.0$). (a)~Complex spectra under the periodic boundary conditions (red dots) and open boundary conditions (black dots). (b)~Complex spectra under the periodic and open boundary conditions in the presence of the flux $\phi \in \{ 0, 2\pi/5, 4\pi/5, 6\pi/5, 8\pi/5 \}$. (c)~Spatial distributions of particle numbers for an eigenstate with $E = -5.98$ under the periodic boundary conditions (red dots) and an eigenstate with $E = -5.60$ under the open boundary conditions (black dots). (d)~Arguments of the determinants of $\hat{H}_{5} \left( \phi \right) - E$ as a function of the flux $\phi$ for $E = 0.0$ (purple dots, $W = 0$), $E=-5.0$ (blue dots, $W=3$), and $E = -9.0$ (yellow dots, $W = 0$).}
	\label{fig: N5}
\end{figure}

Figure~\ref{fig: N5} shows the numerical results for the half-filling case ($L=10$, $N=5$).
We observe the appearance of real eigenenergy whose real part is minimum ($E = -12.1$).
It is $L$-fold degenerate.
Notably, it is insensitive to the flux $\phi$ and robust to the change of the boundary conditions, which is to be contrasted with the isolated loops in the fewer-particle case.
The energy gap between this eigenenergy and the nearest eigenenergy is induced by the many-body interaction similarly to the Mott gap.
If we further increase non-Hermiticity or decrease the interaction, the energy gap decreases;
above a threshold, a phase transition should occur as a consequence of the competition between non-Hermiticity and interactions.
It merits further research to investigate this phase transition, which may be related to the dielectric breakdown of a Mott insulator~\cite{Fukui-98PRB, Oka-10}.

It is also notable that the absence of the skin effect may be a special feature of half-filled ground states (i.e., eigenstates with the minimum real part of energy for the half filling), which is compatible with other non-Hermitian systems~\cite{Lee-20, Liu-20, Alsallom-21}.
While the ground states do not exhibit the skin effect, the open-boundary spectrum for generic eigenstates is different from the periodic-boundary spectrum, which is a clear signature of the skin effect.
Similarly to the fewer-particle cases, the skin modes seem to appear only in the energy regions characterized by the nonzero topological invariant.
In addition, $-\ii \partial_{\phi} \log \det\,[ \hat{H}_1 \left( \phi \right) - E ] = \partial_{\phi}\,\mathrm{arg}\,\det\,[ \hat{H}_1 \left( \phi \right) - E ]$ significantly depends on $\phi$ in contrast to the previous cases.
This behavior may be due to a small system size or a small point gap.

%%%%% Conclusion %%%%%
\section{Discussions}
    \label{sec: conclusion}

We have formulated a topological invariant of non-Hermitian many-body systems in one dimension.
This many-body topological invariant characterizes the winding of the complex spectrum and describes the open quantum dynamics generated by the non-Hermitian operator.
While it reduces to the band topology for noninteracting systems with translation invariance, we have shown that it is free from reduction in the presence of many-body interactions.
As an illustration, we have applied the many-body topological invariant to the interacting Hatano-Nelson model and found the unique complex-spectral winding and concomitant skin effect induced by the interplay of non-Hermiticity and many-body interactions.

In the noninteracting regime, the intrinsic non-Hermitian topological invariant was shown to be the origin of the non-Hermitian skin effect~\cite{Zhang-20, OKSS-20}.
However, the proofs are strongly based on band theory and Hermitization, both of which are no longer applicable in the presence of many-body interactions.
Thus, it is important to revisit the relationship between the topological invariant and the skin effect in non-Hermitian many-body systems.
Moreover, generalizations to other symmetry classes and higher dimensions are a future issue.
It is also of interest to apply the intrinsic non-Hermitian topological invariant to the open quantum dynamics of Liouvillians in master equations~\cite{Diehl-11, *Bardyn-13, Budich-15, *Bardyn-18, Gong-17, Lieu-20, Tonielli-20, Altland-21}.
In particular, it should be relevant to the Liouvillian skin effect~\cite{Song-19, Haga-21, Liu-20PRR, Mori-20}.

%%%%% Note %%%%%
\medskip
{\it Note added.\,---\,}After the completion of this work, we became aware of a recent related work~\cite{Zhang-Neupert-22}.

%%%%% Acknowledgement %%%%%
\section*{Acknowledgements}
We thank Hosho Katsura for helpful discussions.
K.K. thanks Zongping Gong for pointing out the simplified formula in Eq.~(\ref{eq: W - simplified}) for noninteracting systems.
K.K. is supported by KAKENHI Grant No.~JP19J21927 from the Japan Society for the Promotion of Science (JSPS).
K.S. is supported by JST CREST Grant No.~JPMJCR19T2 and JST PRESTO Grant No.~JPMJPR18L4.
S.R. is supported by the National Science Foundation under award number DMR-2001181, and by a Simons Investigator Grant from the Simons Foundation (Award Number: 566116).

\bibliography{NH-topo}

\end{document}